\documentclass[apj]{aeb_emulateapj}
\usepackage{amsmath}
\usepackage{amssymb}

\usepackage{natbib}
\def\s{{\rm s}} 
\def\Ms{{\rm M}\s} 

\def\Hz{{\rm Hz}} 
\def\kHz{{\rm kHz}} 
\def\MHz{{\rm MHz}} 
\def\GHz{{\rm GHz}} 

\def\m{{\rm m}} 
\def\mm{{\rm m}\m} 
\def\cm{{\rm c}\m} 
\def\km{{\rm k}\m} 
\def\pc{{\rm pc}} 
\def\kpc{{\rm k}\pc} 

\def\Ms{M_\odot} 

\def\eV{{\rm eV}} 
\def\erg{{\rm erg}} 

\def\G{{\rm G}} 
\def\muG{\mu\G} 
\def\mG{{\rm m}\G} 

\def\mas{{\rm mas}} 
\def\rad{{\rm rad}} 

\def\cf{{cf.}} 
\def\eg{{e.g.}} 
\def\ie{{i.e.}} 

\def\ptrs{Phil.~Trans.~Roy.~Soc.} 

\renewcommand{\d}{d}
\newcommand{\e}{{\rm e}}
\newcommand{\sgn}{{\rm sgn}}
\newcommand{\RM}{{\rm RM}}
\newcommand{\RMA}{{\rm RM_S}}
\newcommand{\RMSA}{{\rm RM_{SA}}}
\newcommand{\nuSA}{\nu_{\rm SA}}
\newcommand{\nuR}{{\nu_{\rm R}}}
\newcommand\bmath[1] {\mbox{\boldmath$\rm #1$}}
\newcommand{\HzRG}{H$z$RG}

\begin{document}

\title{Understanding the Geometry of Astrophysical Magnetic Fields}

\author{
Avery E.~Broderick\altaffilmark{1} \& 
Roger D.~Blandford\altaffilmark{2}
}
\altaffiltext{1}{Canadian Institute for Theoretical Astrophysics, 60 St.~George St., Toronto, ON M5S 3H8, Canada; aeb@cita.utoronto.ca}
\altaffiltext{2}{Kavli Institute for Particle Astrophysics and Cosmology, 2575 Sand Hill Rd., Menlo Park, CA 94309, USA}

\shorttitle{Understanding the Geometry of Astrophysical Magnetic Fields}
\shortauthors{Broderick \& Blandford}

\begin{abstract}
Faraday rotation measurements have provided an invaluable technique with
which to measure the properties of astrophysical
magnetized plasmas.  Unfortunately, typical observations provide
information only about the density-weighted average of the magnetic
field component parallel to the line of sight.  As a result, the
magnetic field geometry along the line of sight, and in many cases
even the location of the rotating material, is poorly constrained.
Frequently, interpretations of Faraday rotation observations are
dependent upon underlying models of the magnetic field being probed
(e.g., uniform, turbulent, equipartition).
However, we show that at sufficiently low frequencies, specifically
below roughly $13(\RM/1\,\rad\,\m^{-2})^{1/4}(B/1\,\G)^{1/2}\,\MHz$,
the character of Faraday rotation changes, entering what we term the
``super-adiabatic regime'' in which the rotation 
measure is proportional to the integrated {\em absolute value} of the
line-of-sight component of the field.  As a consequence, comparing
rotation measures at high frequencies with those in this new
regime provides direct information about the geometry of the
magnetic field along the line of sight.  Furthermore, the
frequency defining the transition to this new regime, $\nuSA$, depends
directly upon the
{\em local} electron density and magnetic field strength where the
magnetic field is perpendicular to the line of sight, allowing the
unambiguous distinction between Faraday rotation within and in front
of the emission region.  Typical values of $\nuSA$ range
from $10\,\kHz$ (below the ionospheric cutoff, but above 
the heliospheric cutoff) to $10\,\GHz$, depending upon the details
of the Faraday rotating environment.  In particular, for resolved AGN,
including the black holes at the center of the Milky Way (Sgr A*) and
M81, $\nuSA$ ranges from roughly $10\,\MHz$ to $10\,\GHz$, and
thus can be probed via existing and up-coming ground-based radio
observatories.
\end{abstract}

\keywords{polarization -- radiative transfer -- radio continuum: general -- magnetic fields -- turbulence -- Galaxy: center }

\maketitle

\section{Introduction} \label{I}
Magnetized plasmas are common in astrophysics, playing central
roles in objects as diverse as galaxy clusters, the sites of star
formation, the solar corona and accreting black holes and their
ultra-relativistic outflows.  Despite their importance, however, there
are few observational tools for assessing the physical conditions
within them.  This is especially true for strongly ionized or
diffuse regions, in which molecular line emission is absent.  In these
the best evidence for significant magnetic fields, beyond theoretical
arguments, is due to Faraday rotation observations of intrinsically
polarized sources.

First discovered in optically active crystals \citep{Fara:1846},
Faraday rotation has subsequently become one of the primary methods
for measuring the strength and geometry of astrophysical magnetic
fields.  This has been primarily via the determination of the rotation
measure, defined in terms of the polarization angle, $\Phi$, by 
\begin{equation}
\RM
=
\frac{\d \Phi}{\d \lambda^2}
\simeq
8.12\times10^5\int n_e \bmath{B}\cdot\d\bmath{\ell}\,\,\rad\,\m^{-2}\,,
\label{eq:FR}
\end{equation}
where $n_e$, $B$ and $\ell$ are measured in $\cm^{-3}$, $\G$ and $\pc$,
respectively.  Thus, $\RM$'s provides a line-of-sight averaged
measurement of the line-of-sight magnetic field, weighted by the local
plasma density.  Measured values for the $\RM$ range from nearly
zero to nearly $10^6\,\rad\,\m^{-2}$.

Rotation measure observations have played a critical role in
determining the 
magnetic field strengths in the centers of galaxy clusters
\citep[e.g.,][]{Ge-Owen:93,Ge-Owen:94},
Galactic magnetic field strength and distribution
\citep[e.g.,][]{Men-Ferr-Han:08,Nout_etal:08},
magnetic field strength and structure within ultra-relativistic black hole jets
\citep{Zava-Tayl:04,Stir-Spec-Cawt-Para:04,Mill_etal:05,Broc_etal:07,Khar_etal:09},
magnetic field strengths near, and accretion rate of, the supermassive black hole at
the center of the Milky Way and other nearby low-luminosity active
galactic nuclei (AGN)
\citep[e.g.,][]{Agol:00,Quat-Gruz:00,Brun-Bowe-Falc:06,Macq_etal:06,Marr_etal:07},
and even the solar corona magnetic field
\citep[e.g.,][]{Manc-Span:00}.

However, there are a number of significant ambiguities in Faraday
rotation studies.  The first is the degeneracy between density and magnetic
field strength (i.e., high densities and weak fields vs. low densities
and strong fields).  The second is the degeneracy between weak
large-scale fields and strong tangled fields.  In light of numerical
simulations of magnetohydrodynamic turbulence, and the observation of
supersonic (though sub-Alfv\'enic) turbulence in Galactic molecular
clouds, these make interpreting measured $\RM$'s highly dependent upon
models for the rotating medium.  Since in many cases it is unclear if
the Faraday rotation is occuring {\em in situ} or in a foreground
region distant from the polarized source, there are large
uncertainties in the our understanding of the physical conditions in
the source.

Here we show that at low frequencies Faraday rotation qualitatively
changes, with the $\RM$ becoming proportional to
$\int n_e|\bmath{B}\cdot\d\bmath{\ell}|$.  Comparisons between
$\RM$'s measured within this ``super-adiabatic'' regime and at
high frequencies can probe the line-of-sight geometry of
the magnetic field\footnote{Analogous phenomena include the ``MSW''
  effect in neutrino astrophysics \citep{Bahc:89}}.  Furthermore, the
frequency at which a given source
transitions between the standard and super-adiabatic regimes depends
upon the {\em local} plasma properties at magnetic field reversals
(when the magnetic field is orthogonal to the line of sight), and can
thus distinguish between different candidates for the site of the
rotating medium.

In Section \ref{sec:FRaAPMP} we review the plasma processes
responsible for Faraday rotation and describe the super-adiabatic
regime.  Section \ref{sec:O} deals with the observational consequences
unique to super-adiabatic Faraday rotation generally, while Sections
\ref{sec:IS}  \& \ref{sec:ISS} concern specific classes of sources in
which ordinary Faraday rotation has been observed.  Our conclusions
and implications are summarized in Section \ref{sec:C}. The
mathematical details of an {\em ab initio} determination of the
radiative transfer regimes is presented in the Appendix.

\section{Faraday Rotation \& Adiabatic Plasma Mode Propagation} \label{sec:FRaAPMP}

Faraday rotation is commonly presented in terms of the different
phase velocities of the circularly polarized eigenmodes of magnetized
plasmas.  Specifically, a linearly polarized, monochromatic, incident wave is
decomposed into the two plasma modes, which subsequently propagate
independently, accruing a net phase difference,
$\Delta\Phi = \int \Delta k \d \ell$ (where $\Delta k(\omega)$ is the
difference between the wave-vectors of the two modes), as a
consequence of the anisotropic nature of magnetized plasmas.  This
phase difference between circularly polarized modes results in a net
rotation of the polarization angle by $\Delta\Phi/2$.  Frequently, this
is extended to the case of elliptical plasma modes, which occurs when
$\bmath{B}$ is nearly orthogonal to the line of sight, or if the
electrons are relativistic, producing Faraday conversion (also know as
Faraday pulsations or generalized Faraday rotation), corresponding to
oscillations between linear and circular polarizations as well as a
rotation of the polarization angle.

Implicit in this description of the plasma processes underlying
Faraday rotation is the assumption that the two plasma eigenmodes do
indeed propagate independently, or ``adiabatically'', i.e., there is
no mode crossing.  In the case of uniform plasmas (constant $n_e$,
$\bmath{B}$) this is true, and the foregoing explanation is complete.
However, if the plasma is rapidly changing the situation is more
complicated.  For example, consider the propagation of the fast mode
near the location of a magnetic field reversal, i.e., where
$\bmath{k}\cdot\bmath{B}$ changes sign.  Initially, the mode is nearly
circularly polarized, with the electric vector rotating in the same
sense as the gyrating electrons.  As the component of the magnetic
field decreases the mode becomes increasingly elliptical, changing
into the linearly-polarized extraordinary mode when
$\bmath{k}\cdot\bmath{B}=0$.  On the other side of the magnetic
reversal the mode again becomes elliptical and eventually circular,
though now with the sense of rotation reversed, following the sense of
gyration of the electrons.  Similarly, the slow mode reverses its
polarization by evolving through the linearly polarized ordinary
mode.  The net Faraday rotation is then dependent upon
$\int n_e\left|\bmath{B}\cdot\d\bmath{\ell}\right|$.  Clearly the
conditions for the applicability of Equation (\ref{eq:FR}) depend
upon adiabaticity being broken at field reversals.  We call this the
``standard'' regime, with the regions where the adiabatic condition
fails referred to as the ``strongly coupled''.  However, the
``super-adiabatic'' regime, where the propagation is adiabatic
throughout field reversals, can sometimes be observed with interesting
consequences.

Since the failure of the adiabatic condition occurs near field
reversals, we might expect that the standard and super-adiabatic
regimes are distinguished by the properties of the eigenmodes when
$\bmath{B}\cdot\bmath{k}=0$.  Specifically, we might expect that when
the eigenmodes change sufficiently slowly at field reversals, the
propagation remains adiabatic.  Near reversals the evolution of the
eigenmodes are dominated by the increase in ellipticity, becoming
significantly elliptical only when $|\theta-\pi/2|\lesssim
Y\equiv\nu_B/\nu$, where $\theta$ is the angle between the magnetic
field and the line of sight and $\nu_B$ is the electron cyclotron
frequency.  Thus, the eigenmodes change substantially on length scales
on the order of $Y (\d\theta/\d\ell)^{-1}$ near reversals.  To ignore
mode conversion we require that this be large in comparison to $\Delta
k^{-1}$, which at reversals is given approximately by $c/\pi\nu X
Y^2$, where $X\equiv\nu_P^2/\nu^2$ in which $\nu_P$ is the electron
plasma frequency \footnote{Equivalently, we require the energy of the
  mixing term associated with the evolution of the underlying
  eigenmodes be less than the energy splitting between the two modes}.
Thus we arrive heuristically at the result that when
$c \d\theta/\d\ell \ll \pi \nu X Y^3$ the propagation occurs in the
super-adiabatic regime.

\begin{figure}
\begin{center}
\includegraphics[width=\columnwidth]{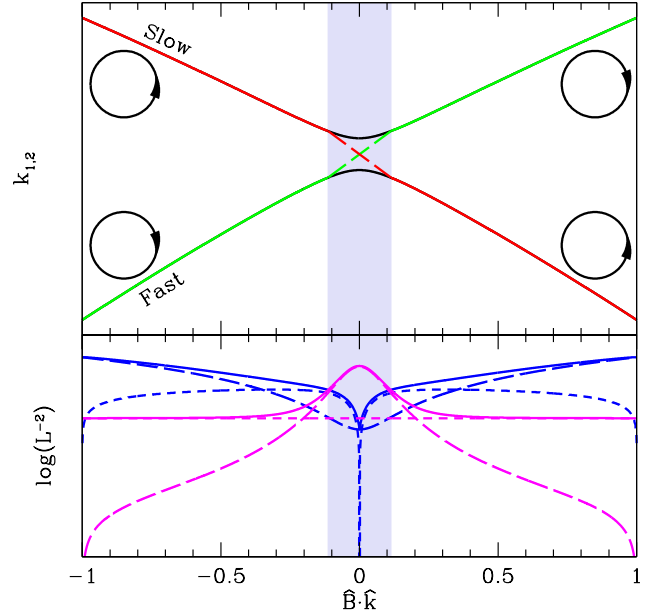}
\caption{Mode behavior near magnetic field reversals.
  {\em Top:}
  The wave numbers of the two plasma eigenmodes, with their
  quasi-longitudinal polarizations shown.  Within the strongly coupled
  regime the amplitudes of the two modes effectively switch, producing
  the standard expression for the rotation measure:
  $\RMA\propto\int n_e \bmath{B}\cdot\d\bmath{\ell}$.
  {\em Bottom:}
  The terms in the adiabatic condition (Equation \ref{eq:ad_cond}).  Blue long
  dash, short dash and solid lines show $(\Delta k/2)^2$,
  $|\Delta k'/2|$ and their sum, respectively.  Magenta long dash,
  short dash and solid lines show $\chi'^2$, $\phi'^2$ and their sum,
  respectively.  When the latter is larger than the former the
  propagation occurs in the strongly coupled regime, denoted by the
  shaded region, and significant mode conversion can occur. }
\label{fig:FR}
\end{center}
\end{figure}

More explicitly, as shown in the Appendix, the necessary
condition for the local plasma modes to propagate adiabatically is
\begin{equation}
\left(\frac{\Delta k}{2}\right)^2
+
\left|\frac{\Delta k'}{2}\right|
\gg
\phi'^2 + \chi'^2\,,
\label{eq:ad_cond}
\end{equation}
where prime denotes differentiation with respect to $\ell$, $\phi$ and
$\chi$ are the orientation and ellipticity angle of the polarization
ellipse, respectively.  However, to second order in the plasma
parameters $\Delta k$ is given by
\begin{equation}
\Delta k = 2\pi\frac{\nu}{c} X Y \sqrt{\cos^2\theta +\left(\frac{Y}{2}\right)^2\sin^2\theta}
\end{equation}
where $\theta$ is the angle between the magnetic field and the line of
sight.  When $\bmath{B}\perp\bmath{k}$, $\Delta k$ is reduced by a
factor of $Y/2$ from its quasi-longitudinal value, and $\Delta k'$
vanishes.  On the other hand, near $\theta=\pi/2$, the eigenmode
ellipticity changes rapidly, giving:
\begin{equation}
\chi'
\simeq
\frac{Y}{4}
\frac{\sin\theta\left(1+\cos^2\theta\right)}{(Y/2)^2+\cos^2\theta}
\theta'\,,
\end{equation}
where we have assumed $Y\ll 1$.  If we further assume that the
direction of rotation of the magnetic field is roughly isotropic,
$\theta'\simeq\phi'$, implying that at reversals $\chi'\simeq \phi'/Y$.
Thus at the same time, the left side of Equation (\ref{eq:ad_cond}) falls
substantially, the right side increases by a similar factor.
This is illustrated explicitly in the bottom panel of
Figure \ref{fig:FR}, which shows the left-side (blue) and right-side
(magenta) of the adiabatic condition through a reversal.  As a
consequence, at magnetic field reversals the mode propagation almost
always becomes strongly coupled, and in the limit of small $Y$ the
polarization propagates as in vacuum.

\begin{figure}
\begin{center}
\includegraphics[width=\columnwidth,bb=18 340 592 718]{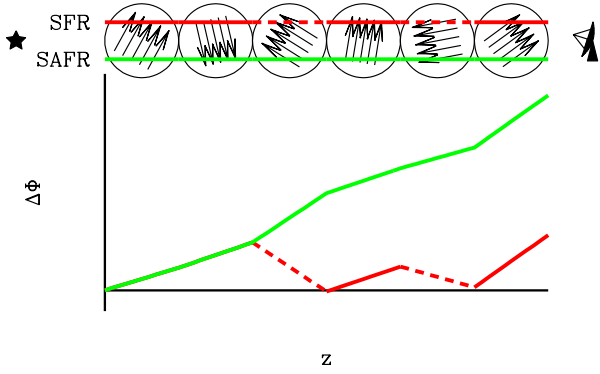}
\caption{The growth the phase difference between the two plasma
  eigenmodes during propagation through a turbulent medium.
  Intervening cells of roughly uniform magnetic field with different
  orientations and strengths are shown on the top, with the source on
  the left and the observer on the right.  The red line shows a path
  taken during standard Faraday rotation, with the phase difference
  accruing during solid segments and decrementing during dashed segments.  The green line
  shows a similar path taken during super-adiabatic Faraday rotation, with
  the phase difference always accruing.  Note the considerable
  enhancement of $\Delta\Phi$ in the super-adiabatic regime, resulting
  in a corresponding enhancement in the rotation measure in this regime.
}
\label{fig:FRvSAFR}
\end{center}
\end{figure}

A direct result of the failure of the adiabatic condition at magnetic
field reversals is that the left-handed (for concreteness) fast mode
maps onto the left-handed slow mode on the other side of the
reversal, and vice versus.  That is, there is an almost complete
conversion of one mode to the other.  Thus, we have the situation
shown by the red line in Figure \ref{fig:FRvSAFR}, as the polarized
wave passes through regions of magnetic field with
$\bmath{B}\cdot\bmath{k}>0$, $\Delta\Phi$ increases (solid lines), while when
$\bmath{B}\cdot\bmath{k}<0$, $\Delta\Phi$ decreases (dashed lines),
yielding the well-known formula,
\begin{equation}
\Delta\Phi \propto \nu^{-2} \int n_e\bmath{B}\cdot\d\bmath{\ell}\,.
\end{equation}
Note that the failure of the adiabatic condition at field reversals
was {\em critical} to obtaining the standard result.

Nevertheless, at sufficiently low frequencies, the adiabatic condition,
Equation (\ref{eq:ad_cond}), can be satisfied at reversals.  That is, because
$|\Delta k'|$ vanishes and $\chi'\gg\phi'$ where
$\bmath{B}\cdot\bmath{k}=0$, in order to remain in the adiabatic mode
propagation regime throughout, what we call the ``super adiabatic''
regime, we require
\begin{equation}
\frac{\theta'}{Y}
\ll
\frac{\pi}{2} \frac{\nu}{c} X Y^2
\quad
\Rightarrow
\quad
\nu \ll \nuSA \equiv \left[ \frac{\pi\nu_B^3\nu_P^2}{2 c \theta'} \right]^{1/4}\,.
\end{equation}
Note that up to a factor of two, this is precisely the condition we
reached earlier by heuristic arguments.  The critical frequency below
which the propagation is super-adiabatic may be recast in terms of the
local magnetic field reversal length scale,
$\ell_B\equiv\pi/2\theta'\equiv10^{15}\ell_{B,15}\,\cm$,
and the local plasma parameters:
\begin{equation}
\nuSA \simeq 87 n^{1/4} B^{3/4} \ell_{B,15}^{1/4}\,\MHz\,,
\end{equation}
in which $n$ and $B$ are measured in $\cm^{-3}$ and $\G$,
respectively.  Alternatively, this may be written in terms of the
total number of reversals along the line of sight, $N\simeq \theta' L/2\pi$,
the rotation measure determined at frequencies above $\nuSA$,
$\RMA\equiv\lambda^{-2}\Delta\Phi \simeq \left.\lambda^{-2}\Delta k\right|_{\theta=0}L /\sqrt{N}$ (where $N$ enters due to the cancellations discussed
previously) and the magnetic field strength:
\begin{equation}
\begin{array}{rcl}
\displaystyle
\nuSA
&\simeq&
\displaystyle
\left(\frac{\nu_Bc}{2\sqrt{2\pi}}\right)^{1/2} \RMA^{1/4} \,N^{-1/8}\\
&&\\
&\simeq&
\displaystyle
13 \,B^{1/2}\, \RMA^{1/4}\, N^{-1/8}
\,\MHz\,,
\end{array}
\label{eq:nuSA}
\end{equation}
where in the final expression $B$ and $\RMA$ are in  $\G$ and
$\rad\,\m^{-2}$, respectively.  Note that this is only a weak function
of $\RM_S$ and $N$, and thus primarily indicative of the local magnetic
field strength in the source.  As we shall see in Section \ref{sec:IS}, for
a number of sources this produces $\nuSA$ near $100\,\MHz$--$1\,\GHz$.

\begin{figure}
\begin{center}
\includegraphics[width=\columnwidth]{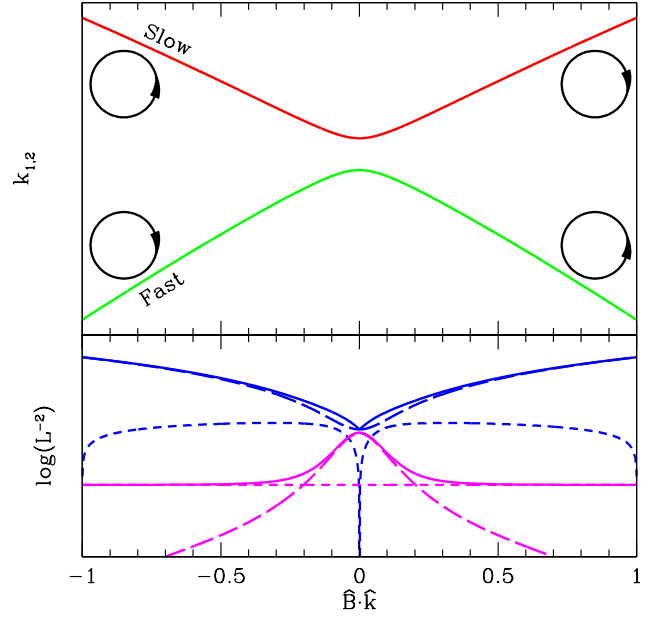}
\caption{Mode behavior near magnetic field reversals in the
  super-adiabatic regime.  {\em Top:}
  Same as in Figure \ref{fig:FR}, except now the modes are no longer
  coupled near $\bmath{B}\cdot\bmath{k}=0$.  As a result, the fast/slow
  mode remains the fast/slow mode, though the local polarizations
  reverse, and $\RMSA\propto \int n_e |\bmath{B}\cdot\d\bmath{\ell}|$.
  {\em Bottom:} Same as in Figure \ref{fig:FR}, though note the
  absence of a strongly coupled regime.
}
\label{fig:SAFR}
\end{center}
\end{figure}

Since in the super-adiabatic regime there is no mode crossing at
magnetic field reversals, the fast/slow mode maps onto the fast/slow
mode, as illustrated in Figure \ref{fig:SAFR}.  As a result, the
phase difference between the modes only accrues, as seen by the green
line in Figure \ref{fig:FRvSAFR}.  That is, for super-adiabatic
Faraday rotation, 
\begin{equation}
\Delta\Phi_{\rm SA}
\propto
\nu^{-2} \int n_e \left| \bmath{B}\cdot\d\bmath{\ell} \right|\,.
\end{equation}

\newpage
\section{Observational Consequences of Super-Adiabatic Faraday Rotation}\label{sec:O}

There are three additional fundamental constraints upon observable values of
$\nuSA$.  The first is the requirement that within the Faraday screen,
$\nuSA$ must exceed the upper cutoff of the extraordinary mode:
\begin{equation}
\nuSA>\frac{\nu_B}{2} \left(1+\sqrt{1+4\frac{\nu_P^2}{\nu_B^2}}\right)\,.
\end{equation}
Inserting the Equation (\ref{eq:nuSA}) and assuming that there is
considerable separation between the cyclotron and plasma scale gives two
conditions upon the magnetic field,
\begin{equation}
B > 2.1\times10^{-5}\,\RMA^{1/4} N^{-1/8} \ell_{B,15}^{1/2}\,\G\,,
\label{eq:nuSAnuB}
\end{equation}
corresponding to $\nuSA>\nu_P$, and
\begin{equation}
B < 22\, \RMA^{1/2}\, N^{-1/4}\,\G\,,
\label{eq:nuSAnuP}
\end{equation}
corresponding to $\nuSA>\nu_B$\footnote{This also implies that we may
  ignore Razin suppression.}.
Note that this produces lower {\em and} upper and bound upon $B$ as a
function of $\RMA$.  The former is shown for various values of
$N^{-1/4}\ell_B$ by the dashed lines in Figure \ref{fig:regimes}.  The
latter is denoted by the upper-left greyed region in Figure
\ref{fig:regimes}.

The second is that we must be able to ignore difference in the
refraction of the two plasma eigenmodes.  As the plasma eigenmodes
propagate through the Faraday screen they necessarily follow different
paths, dictated by their different indices of refraction.  If the two
paths pass through sufficiently distinct regions of the Faraday
screen, Equation (\ref{eq:FR}) is no longer valid, $\Phi$ depending
upon the differences in the plasma conditions along the paths in
addition to the difference in the phase velocities of the two modes,
and the $\RM$ is no longer constant.  As shown in the Appendix,
the condition that the modes propagate through strongly correlated
plasma regions is $\nu\gg\nuR$ where
\begin{equation}
\begin{array}{rcl}
\nuR
&\simeq&
\displaystyle
61 \frac{D^{1/3} L^{1/6}}{\ell_B^{1/2}} n^{1/3} B^{1/3} f^{1/3}\,\kHz\\
&&\\
&\simeq&
\displaystyle
2.0 D_{22}^{1/3} \ell_{B,15}^{-2/3} \RM^{1/3} f^{1/3}\,\MHz
\end{array}
\label{eq:refrac}
\end{equation}
in which $f$ is the fractional variation in the plasma parameters,
$D=\left(D_O^{-1}+D_S^{-1}\right)^{-1}=D_{22}10^{22}\,\cm$ where $D_O$
and $D_S$ are the observer--screen and source--screen distances,
respectively.  The nature of this constraint depends upon the location
of the Faraday screen.  For example, when the rotation occurs
{\em in situ}, $D\simeq L$ implying that $D/\ell_B\simeq N$, and
giving $\nu_R$ that is a moderately weak function of $N$ ($\propto
N^{1/2}$ or $N^{1/3}$ depending upon which of the above expressions
one considers).  For all of the sources we describe below,
$\nuSA>\nuR$, justifying our neglect of refraction for these objects.

The third is the observational limits due to absorption in the
ionosphere and heliopause \citep[see, e.g.,][]{Boug:96}.  The
ionosphere limits ground observations to above roughly $15\,\MHz$,
though down to $10\,\MHz$ is possible at excellent sites.  A
fundamental limit for space missions is the plasma frequency at the
heliopause, estimated at roughly $9\,\kHz$.  These are shown by the
hatched and solid grey regions, respectively, in the lower left.
While a number of interesting objects lie within reach of ground-based
radio astronomy, including nearby AGN and pulsars, pushing to
space-based, ultra-low frequency radio observatories opens a new window
upon a vast array of astrophysical environments.

If it can be observed, Faraday rotation in the super-adiabatic regime,
what we term ``super-adiabatic Faraday rotation'', provides a wealth of new
information about the magnitude and line-of-sight geometry of magnetic
fields.  First, unlike standard Faraday rotation the super-adiabatic
rotation measure:
\begin{equation}
\RMSA
=
C \int n_e \left|\bmath{B}\cdot\d\bmath{\ell}\right|\,,
\end{equation}
does not suffer from a degeneracy between weak large-scale magnetic
fields and strong tangled magnetic fields.  Thus, $\RMSA$ is far
better suited for the determination of $n_e$ and $B$ via equipartition
arguments than its high-frequency analog:
\begin{equation}
\RMA
=
C \int n_e \bmath{B}\cdot\d\bmath{\ell}\,.
\end{equation}

Second, generally, $\RMSA$ will be greater than $\RMA$ by an amount
depending upon the number of magnetic field reversals along the line
of sight.  Typically, we may expect that looking through $N$
independent turbulent cells, the excess phase difference, and
therefore $\RMA$, to scale as $N^{-1/2}$, i.e.,
\begin{equation}
\RMSA \simeq N^{1/2} \RMA\,.
\end{equation}
Thus, measuring both $\RMA$ (at $\nu>\nuSA$) and $\RMSA$ (at
$\nu<\nuSA$) provides an estimate for the number of magnetic
field reversals along the line of sight:
\begin{equation}
N \simeq \left(\frac{\RMSA}{\RMA}\right)^2\,.
\end{equation}
However, this estimate may not be appropriate in the presence of
long-range order in the magnetic field, and more than a rough estimate
requires a detailed analysis of the nature of the magnetic fields
under consideration.  For example, if there was some reason to not
expect a randomly reversing field, but instead a band-structured field
(e.g., as might be expected from pulsars or hydromagnetic turbulence
within a given cell), then, $\RMSA/\RMA\simeq N$.

Third, as seen in Equation (\ref{eq:nuSA}), the location of the
super-adiabatic transition depends upon the {\em local} parameters of
the plasma, though we have recast them in terms of the known $\RMA$
and $B$ by assuming that the intervening Faraday screen isn't
significantly clumpy.  However, in many sources this assumption may
not be appropriate.  Measuring $\nuSA$ provides a {\em direct}
measurement of the {\em local} plasma properties at the magnetic field
reversals, and thus the distribution of the electron density and
magnetic field strength along the line of sight.  On the other hand,
if we can assume that these quantities are smoothly distributed,
measuring $\RMSA$ and $\nuSA$ provides a method to estimate the
magnetic field strength, independent of the intervening electron
density:
\begin{equation}
B \simeq
6
\,\nuSA^2
\left(\frac{\RMSA}{\RMA}\right)^{1/4}
\RMA^{-1/2}
\,\mG\,,
\end{equation}
where $\nuSA$ and $\RMA$ are measured in $\MHz$ and $\rad\,\m^{-2}$,
respectively.

However, both bandwidth and beam depolarization are likely to be
more severe problems for super-adiabatic Faraday rotation than its
standard counterpart.  Both are exacerbated simply because
$\RMSA>\RMA$.  Beam depolarization, however, is also more likely due
to the nature of super-adiabatic propagation.  At field reversals the
polarization is effectively reflected about the magnetic field as the
handedness of the polarization eigenmode switches.  The observed
polarization is therefore dependent in part upon the magnetic field
direction in the last turbulent cell.  This can produce large
gradients in the polarization position angle on angular scales
comparable to the turbulent scale.  Furthermore, because probing the
super-adiabatic regime requires pushing to longer wavelengths, it also
generically results in larger beams.  Nevertheless, for compact
sources, those that subtend a small angular scale in comparison to the
turbulence in the intervening Faraday screen, this effect is largely
mitigated.

\section{Implications for Specific Sources:\\ Ground Based Observations}\label{sec:IS}

\begin{figure*}
\begin{center}
\includegraphics[width=\textwidth]{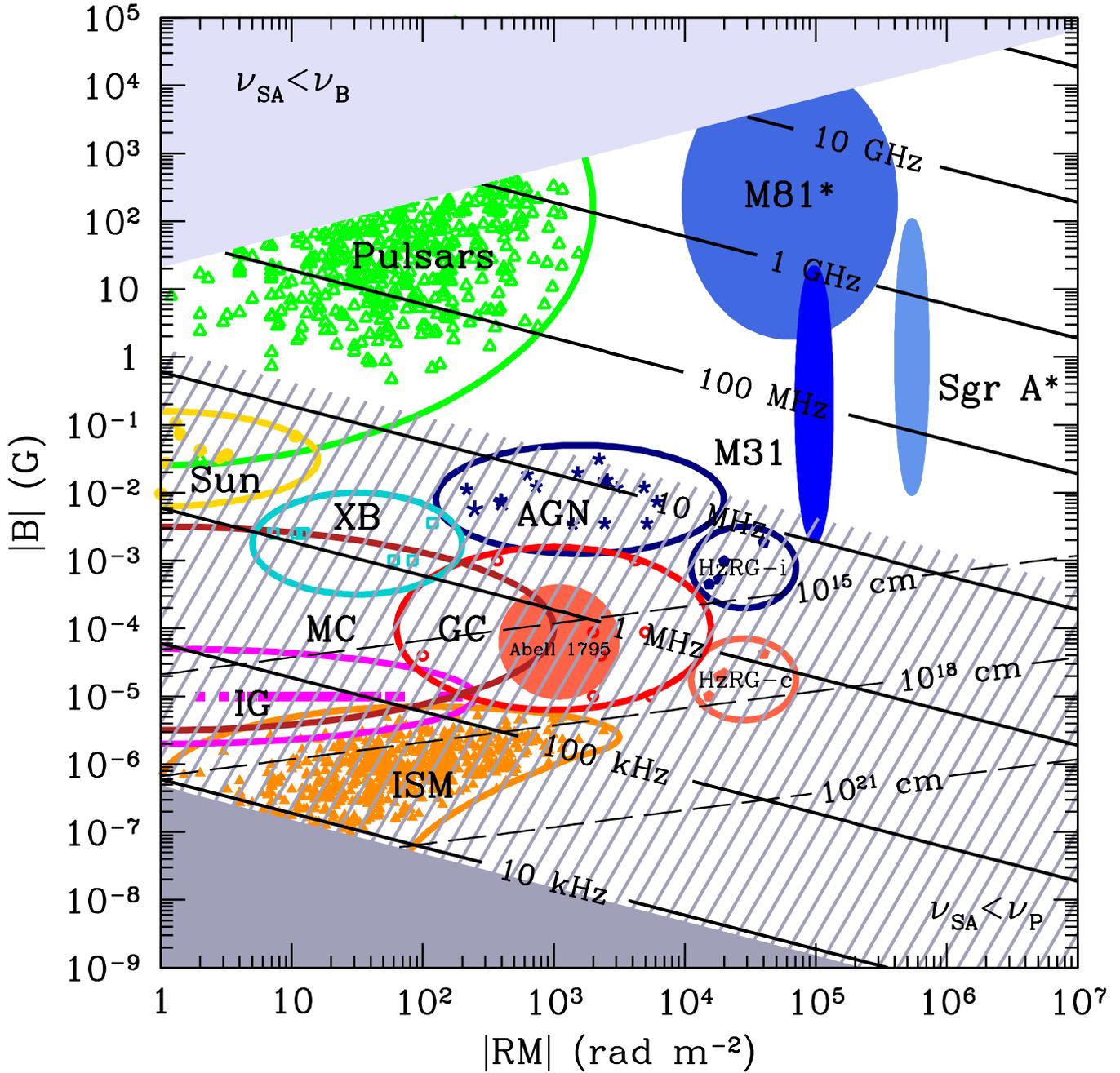}
\caption{Lines of constant $\nuSA$ for various rotation measures and
  source magnetic field strengths, assuming $N=1$ (though since
  $\nuSA$ is so weakly dependent upon $N$ it is unlikely to be
  significantly different in practice).  We require that $\nuSA$ lie
  above the upper-cutoff frequency, which reduces to roughly the
  condition that $\nuSA>\nu_B$ and $\nuSA>\nu_P$ (see Equations
  \ref{eq:nuSAnuB} \& \ref{eq:nuSAnuP}).  The first of these
  corresponds to $B<22\RMA^{1/2}N^{-1/4}\,\G$, and is
  violated within the upper-left grayed region.  The second reduces to
  $B>21\RMA^{1/4}N^{-1/8}\ell_B^{1/2}\,\muG$ and is shown by the thin
  dashed lines in the lower-right for $\ell_B=10^{15}\,\cm$,
  $10^{18}\,\cm$ and $10^{21}\,\cm$. The lower-left grey-hatched
  region denotes where $\nuSA$ lies below
  the ionospheric cutoff ($\sim 15\,\GHz$) and the low-left greyed region
  denotes where $\nuSA$ lies below the heliospheric cutoff ($\sim
  10\,\kHz$).  The locations, and therefore the expected $\nuSA$, for
  various classes of potential sources are shown, based upon observed
  $\RM$'s and estimated magnetic field strengths.  Points and filled
  ellipses correspond to individual sources, while the empty regions
  are rough estimates of the envelope bounding the region each class
  of sources occupies.  The represented source classes, and the sections
  discussing them in more detail, are:
  Sgr A* (\ref{sec:RGN:SgrA}),
  M81 (\ref{sec:RGN:M81}),
  M31 (\ref{sec:RGN:M31}),
  AGN cores \& jets and high-redshift radio galaxies (AGN \& \HzRG-i, \ref{sec:UAGN}),
  intrinsic rotation in Pulsars (Pulsars, \ref{sec:Pi}),
  the solar corona (Sun, \ref{sec:SC}),
  the interstellar medium (ISM, \ref{sec:PISM}),
  X-ray binaries (XB, \ref{sec:XB}),
  Galactic center region (GC, \ref{sec:GC}),
  Galactic molecular clouds (MC, \ref{sec:GMC}),
  infrared/starburst galaxies (IG, \ref{sec:SG})
  and
  the intracluster medium (Abell 1795 \& \HzRG-c, \ref{sec:ICM}).
}
\label{fig:regimes}
\end{center}
\end{figure*}

We now turn to the implications for specific classes of sources, and
in particular estimate $\nuSA$.  Figure \ref{fig:regimes} summarizes
these.  Here we dicuss those sources for which the super-adiabatic
regime is accessible from ground based observations, i.e.,
$\nuSA\gtrsim15\,\MHz$.

\subsection{Ionospheric Faraday Rotation}\label{sec:IFR}
The ionosphere presents an opportunity to investigate the
super-adiabatic regime via local radio experiments.
The F-layer is roughly $300\,\km$ high with a typical
free-electron density of $10^6\,\cm^{-3}$.  Using a typical
terrestrial magnetic field strength of $0.5\,\mG$, this results in an
$\RM$ of roughly $4\,\rad\,\m^{-2}$ at zenith, and potentially
considerably more at oblique angles.  This will be further enhanced at
low frequencies by ionospheric refraction.  Thus, while
moderate in comparison to most astronomical sources, the atmospheric
rotation measure is certainly detectable.

Near the poles the Earth's magnetic field necessarily changes
direction, with $\theta'\simeq\pi/R_\oplus$.  Thus,
$\nuSA\simeq40\,\MHz$, which is well above the ionospheric cut-off.
While ionospheric refraction is significant at this frequency, the
differential refraction is sufficiently small ($\nu_R\simeq5\,\MHz$).
Thus, transpolar observations of the Jovian or Saturnian decametric
radiation provides a natural terrestrial radio experiment in which to
study the transition from the standard to the super-adiabatic regimes
in a well understood plasma.

\subsection{Resolved Galactic Nuclei} \label{sec:RGN}

\subsubsection{Sagittarius A*} \label{sec:RGN:SgrA}
The supermassive black hole at the center of the Milky Way,
Sagittarius A* (Sgr A*), is presently the best studied accreting black
hole.  Despite its comparatively small mass
\citep[$4.7\times10^6\,\Ms$,][]{Ghez_etal:08,Gill_etal:09} and
extraordinarily low luminosity ($L/L_{\rm Edd}\simeq10^{-10}$), due
to its proximity it has the unique distinction of being the only
black hole to have been probed on sub-horizon scales
\citep{Doel_etal:08}.  It also has the largest rotation measure
observed, $-5.6\pm0.7\times10^5\,\rad\,\m^{-2}$
\citep{Marr_etal:07,Macq_etal:06}.  Estimates of Sgr A*'s magnetic
field arise from both fitting simple accretion models to both the
rotation measure results and the source spectrum
\citep{Agol:00,Quat-Gruz:00,Falc-Mark:00,Yuan-Quat-Nara:03,Marr_etal:07,Loeb-Waxm:07},
and range from $0.01$--$100\,\G$.  The considerable uncertainty
reflects the uncertainty in the location region responsible for the
Faraday rotation as well as the geometry of the magnetic field along
the line of sight.  Nevertheless, these estimates imply that Sgr A*
should enter the super-adiabatic regime between $30\,\MHz$ and
$3\,\GHz$.

Typical length scales within the Faraday rotating medium surrounding
Sgr A* are on the order of $\ell_B\simeq10^3GM/c^2\simeq
10^{15}\,\cm$.  Even with the large $\RM$'s observed in Sgr A*, this
still results in a $\nuR\simeq1\,\MHz$, and thus refractive decoherence
is unimportant.

\subsubsection{M81*} \label{sec:RGN:M81}
The supermassive black hole at the center of M81 is a second
representative of a class of low-luminosity AGN.  At
$7\times10^6\,\Ms$, it is moderately more massive that Sgr A*, and
with $L/L_{\rm Edd}\simeq 10^{-5}$, considerably brighter, though
still considerably sub-Eddington \citep{Brun-Bowe-Falc:06}.  The lack
of polarization in M81 at wavelengths above $3.6\,\mm$ implies lower
limits upon the value of $\RM$ of $10^4\,\rad\,\m^{-2}$ 
and $4\times10^5\,\rad\,\m^{-2}$, depending upon whether the source is
beam or band depolarized \citep{Brun-Bowe-Falc:06}.  Estimates for the
magnetic field in the Faraday screen are again made by appealing to
modeling the spectrum of M81* and in analogy with Sgr A* (see above),
giving a range of roughly $2$--$2\times10^4\,\G$, due to its larger
luminosity and mass.  This corresponds to somewhat larger $\nuSA$,
near $3\,\GHz$, though closer to the $\nuSA\simeq\nu_B$ cutoff.  Again
refraction is unimportant in this source.

\subsubsection{Other Resolved AGN} \label{sec:RGN:M31}
Determining the magnetic field strengths within the Faraday
screen of nearby AGN other than Sgr A* and M81* is the primary
difficulty in assessing the likelihood of observing super-adiabatic
Faraday rotation in these sources.  For other resolved, sub-Eddington
AGN we may estimate the 
magnitude of the magnetic field in the accretion flow via scaling
arguments, equipartition and Sgr A*.  Specifically, in terms of the
luminosity in Eddington units,
$\ell_{\rm Edd} = L/L_{\rm Edd} \propto \dot{M}/\dot{M}_{\rm Edd}$,
the electron (and therefore ion) density near the black hole scales as 
\begin{equation}
n_e 
\propto
\frac{\dot{M}}{M^2 c}
\propto
\ell_{\rm Edd} M^{-2} \dot{M}_{Edd}
\propto
\ell_{\rm Edd} M^{-1}\,,
\end{equation}
where we have assumed that the radiative efficiency is not a strong
function of mass for substantially sub-Eddington sources.  The ion
temperature is largely independent of black hole mass, reaching
$10^{11}\,{\rm K}$ in nearly all cases.  Therefore, the equipartition
magnetic field strength scales as
\begin{equation}
B \propto \left( n_e k T \right)^{1/2} \propto \ell_{\rm Edd}^{1/2} M^{-1/2}\,.
\end{equation}
We can also use this to estimate the expected scaling of rotation measures:
\begin{equation}
\RM \propto n_e B d\ell \propto \ell_{\rm Edd}^{3/2} M^{-1/2}\,.
\label{eq:RMacc}
\end{equation}
Note that these imply that brighter and lower-mass sources will have
higher values of $\nuSA$.  Since the rotation occurs locally in these
sources, $\nuSA/\nuR\propto\ell_{\rm Edd}^{1/2} M^{1/4}$, and thus
refractive decoherence is unlikely to ever preclude observing the
super-adiabatic regime for resolved AGN.  

The black hole at the center of M31 is shown as an example.  Despite
being much more massive, $M\simeq1.4\times10^8\,\Ms$, it also is
extraordinarily under-luminous, with $\ell_{\rm Edd}\simeq 10^{-10}$,
similar to Sgr A*~\citep{Li-Wang-Wakk:09}.  The correspondingly
rescaled $\RM$ and $B$ is
shown in Figure \ref{fig:regimes}.  More generally, polarization
observations of resolved nuclear regions will fill the upper-right
corner of the $B$--$\RM$ plane, corresponding to typical $\nuSA$
ranging from the ground-based cutoff of $15\,\MHz$ to $10\,\GHz$.

\subsection{Unresolved AGN}\label{sec:UAGN}
Faraday rotation has been one of the primary methods for studying the
magnetic geometry of AGN radio jets and cores.  $\RM$ gradients have
even produced direct evidence for a helical field geometry in AGN jets
\citep[see, e.g.][]{Khar_etal:09}.  Presently, rotation measures
exist for more than 40 AGN, ranging from roughly $10^2\,\rad\,\m^{-2}$
to nearly $10^4\,\rad\,\m^{-2}$ \citep{Zava-Tayl:04}.  These are
usually determined via fitting VLBA observations at $8\,\GHz$ and
$15\,\GHz$, and in nearly all cases does not resolve the central
source.  In principle, these are associated with the accreting black
hole or jet, and thus are in many respects similar to the systems
discussed in Section \ref{sec:RGN}, though generally with considerably
higher $\ell_{\rm Edd}$.

Estimating the magnetic field is again the primary challenge.  We
obtain an approximate magnetic field strength based upon an
equipartition argument with the emitting nonthermal electrons, assumed
to be accelerated in relativistic shocks either in the accretion flow
or in the jet.  In this case we expect the electron power-law index to
be roughly $-2$ between some minimum and maximum Lorentz factors, and
$L_\nu$ is given locally by 
\begin{equation}
L_\nu \simeq 3\times10^{32}
\gamma_{\rm min}
n_e
B^{3/2}
V
\nu^{-1/2}
\,\erg\,\s^{-1}\Hz^{-1}\,,
\label{eq:AGN_L}
\end{equation}
where $n_e$, $B$, $V$ and $\nu$ are given in $\cm^{-3}$, $\G$, $\pc^3$
and $\GHz$, respectively \citep[the flat spectrum of AGN are then produced
by spatial structure in these quantities, see, e.g.,][]{Blan-Koni:79}.
The electron pressure is given by
\begin{equation}
P_e
\simeq
\frac{1}{3} \epsilon_e
=
\frac{1}{3} \gamma_{\rm min} m_e c^2 n_e
\ln\left(\frac{\gamma_{\rm max}}{\gamma_{\rm min}}\right)
\label{eq:AGN_P}
\end{equation}
If the ion pressure is similar, equipartition gives $P_e\simeq\beta
B^2/8\pi$, for typical $\beta$'s of $10$--$100$.  Thus, upon combining
Equations (\ref{eq:AGN_L}) \& (\ref{eq:AGN_P}), given $\nu$, $L_\nu$, and
$V$, we obtain a lower limit upon the field strength:
\begin{equation}
B
\simeq
2\times10^2\,
\left(\Lambda/\beta\right)^{-2/7} L_{\nu,35}^{2/7}\, \nu^{1/7} V^{-2/7}\,\mG\,,
\label{eq:Beq}
\end{equation}
where $L_\nu = 10^{35} L_{\nu,35}\,\erg\,\s^{-1}\Hz^{-1}$,
$\Lambda\equiv\ln(\gamma_{\rm max}/\gamma_{\rm min})$ and the
other quantities are given in the units described previously.  Note
that this is weakly dependent upon $\beta$, and thus the
system would need to be extraordinarily sub-equipartition for the
magnetic field to be appreciably lower.  For concreteness we choose
$\Lambda\simeq10$ and $\beta\simeq10^2$.

For the AGN jets \& cores for which we found $\RM$'s in the
literature \citep{Zava-Tayl:04,Khar_etal:09}, at $\nu=15\,\GHz$,
$L_{\nu,35}$ is typically of order unity.  More difficult is
estimating $V$, since the vast majority of AGN cores are
unresolved, and we can therefore only place an upper limit upon $V$
(lower limit upon the equipartition value of $B$).  To estimate
this, for each AGN we inspected high-frequency VLBI images to
determine approximate upper limits upon its angular size, typically
comparable to beam sizes of $\sim 1\,\mas$
\citep{Tayl:98,List-Mars-Gear:98,Tayl:00,List-Smit:00,List:01,Porc-Rioj:02,Zava-Tayl:03,Zava-Tayl:04,Jors_etal:05,List_etal:09}.  Given the distances inferred from
the source redshifts, this corresponds to limits upon the physical
sizes of roughly $5\,\pc$, or volumes of $10^2\,\pc^3$--$10^3\,\pc^3$.
The resulting estimates for the magnetic field strength are
$3\,\mG$--$30\,\mG$, and shown by the dark-blue region labeled ``AGN''
in Figure \ref{fig:regimes}.

In addition to the AGN discussed above, there are a small number of
high-redshift radio galaxies (\HzRG) exhibiting anomalously large
rest-frame rotation measures.  These include
OQ 172 \citep[$z=3.535$, $\RM\simeq4\times10^4\,\rad\,\m^{-2}$,][]{Udom-Tayl-Pear-Robe:97}, 
3C 295 \citep[$z=3.377$, $\RM\simeq2\times10^4\,\rad\,\m^{-2}$,][]{Perl-Tayl:91,Tayl-Perl:92}, 
SDSS 1624+3758 \citep[$z=0.461$, $\RM\simeq1,8\times10^4\,\rad\,\m^{-2}$,][]{Frey_etal:08},
and
PKS B0529-549 \citep[$z=2.575$, $\RM\simeq1.5\times10^4\,\rad\,\m^{-2}$,][]{JWBrod_etal:07}.
These $\RM$ are an order of magnitude larger than those discussed
previously, and thus may represent a distinct class.
Their association with cooling-core clusters has led to speculation
that the Faraday rotation is associated with the intracluster medium
and not intrinsic to the broad-line region.  Nevertheless, if the
rotation does occur within the core, this implies typical magnetic
fields of $0.5\,\mG$, somewhat lower than those for other AGN due to
their comparatively larger distances.  This is shown in the dark-blue
region labeled ``\HzRG-i''.

Resolved AGN exhibit strong $\RM$ gradients, frequently with
$\RM$ dispersions that exceed the averaged $\RM$.
As a result, the $\RM$ for unresolved AGN cores is likely to be
considerably smaller than the peak values.  Furthermore,  as we have
stressed above, our estimates of the local magnetic field are highly
uncertain.  This is primarily a consequence of the unresolved natures
of most AGN, implying that our estimate of the equipartition magnetic
field is only a lower limit.  If the Faraday rotation is occuring near
the supermassive black hole, as appears to be the case in Sgr A*, then
the AGN region should move up and to the right.  Therefore, while for
AGN typical $\nuSA$'s presently lie just at the frequency cutoff for
ground-based observations, in reality it may be closer to those for
their resolved counterparts.  Furthermore, for {\em in situ} rotation,
$\nuR\simeq 0.3 N^{1/2} \ell_{B,18}^{-1/3} \RM_3^{1/3}\,\MHz$, still
comfortably below $\nuSA$.

On the other hand, if the Faraday screen
is far from the region responsible for the emission, the intrinsic
magnetic field may be closer to $1\,\muG$--$100\,\muG$, typical of the
interstellar medium.  In this case, the $\nuSA$'s would lie well
below the ground-based cutoff, though above the heliospheric cutoff,
implying that in these sources super-adiabatic Faraday rotation may be
observable from ultra-low frequency, space-based radio observatories.
Thus, even just the detection of super-adiabatic Faraday rotation
would determine the region responsible for the observed Faraday
rotation.

\subsection{Pulsars: Intrinsic Rotation} \label{sec:Pi}
The rotation measures of pulsars can be quite large, reaching
$3\times10^3\,\rad\,\m^{-2}$ in some sources.  
Traditionally, this has been interpreted to be due to Faraday rotation
in the interstellar medium (ISM, see Section \ref{sec:PISM}).
Addressing the propagation of radio waves through pulsar
magnetospheres has been studied extensively elsewhere
\citep[see, e.g.,][]{Aron-Barn:86,Barn-Aron:86,Thom-Blan-Evan-Phin:94},
and is
beyond the scope of this paper.  Generally, it has been
found that the propagation modes are quite different than those we
have described here, due to structure within the magnetosphere and the
presence of a dominant population of relativistic particles.
Nevertheless, it is not clear that propagation through the plerion
does not result in a substantial contribution to the observed
$\RM$'s.

Within the plerion, the wind is again dominated by relativistic
particles, which have strongly elliptical polarization modes,
complicating the determination of the expected $\RM$'s.  In addition,
the wind has a relativistic bulk motion, which can significantly
alter Equation (\ref{eq:FR}), \citep[see, e.g.,][]{Brod-Loeb:09}.  For
objects like the Crab, for which the bulk Lorentz factor can reach
$10^6$, these relativistic effects can dominate the Faraday rotation,
depending upon where the Faraday rotating medium is located.

Given these caveats, observing super-adiabatic Faraday rotation in
pulsars would present direct evidence for intrinsic Faraday rotation in
pulsars, with the attendant consequences for the geometry of pulsar
magnetospheres and recalibrating Galactic magnetic field studies.
Given the uncertainties mentioned above, it is unclear what the
relevant magnetic field strength is in this case.  Here we take as a
fiducial value that at the light cylinder,
$B_{\rm lc}\simeq B_{\rm surf}(R_{\rm NS}/R_{\rm lc})^3 = (2\pi R_{\rm NS}/P c)^3$, 
typically on the order of $10\,\G$--$10^3\,\G$, though in some cases
considerably higher.  While the magnetic field strength is a strong function of
radius (though a considerably weaker function in the wind region,
where $B$ is roughly proportional to $r^{-1}$, than
in the magnetosphere in which $B\propto r^{-3}$), it does not reach
interstellar values until roughly $3\times10^{-6}\,\pc$, vastly larger
than the light cylinder.
If the entire measured $\RM$ is due to the region near the pulsar, the
implied magnetic field strengths would place pulsars in the upper-left
of Figure \ref{fig:regimes} for 749 objects, collected from
\citet{Nout_etal:08} and the ANTF catalog at
www.atnf.csiro.au/research/pulsar/psrcat
\citep{Manc-Hobb-Teoh-Hobb:05}, shown in green and labeled ``Pulsars''.

\section{Implications for Specific Sources:\\ Space Based Observations}\label{sec:ISS}

Despite lying below the ionospheric cutoff, the super-adiabatic regime
for many objects is potentially accessible from space
\citep[e.g.,][]{Jone_etal:00}.  These necessarily have $\nuSA\gtrsim
9\,\kHz$, and are thus above the heliospheric cutoff.  However, in
many cases they are considerably closer to their respective refractive
decoherence limits. Nevertheless, the observation of the
super-adiabatic regime provides a prime motivation to push towards a
space-based, low-frequency radio emission.

\subsection{Solar Corona} \label{sec:SC}
Faraday rotation measurements of polarized radio sources near the
solar disk provide a unique method with which to probe the magnetic
field of the solar corona.  Within this region the solar magnetic
field is still quite strong, ranging from $10\,\mG$ to $0.2\G$, and
dense, with an electron density on the order of $10^3\,\cm^{-3}$.
Nevertheless, the short path length results in only modest $\RM$'s,
ranging between $10\,\rad\,\m^{-2}$ and zero \citep{Manc-Span:00}.
These imply that $\nuSA\simeq3\,\MHz$, shown by the yellow region
labeled ``Sun'' in Figure \ref{fig:regimes}.  The refractive limit is
roughly $0.4 D_{11}^{1/3} \ell_{B,11}^{-2/3} \RM_1^{1/3}\,\MHz$, and
thus roughly an order of magnitude below this.  Therefore, potential
low-frequency, space-based radio observatories could directly probe
the Corona for large-amplitude, long-wavelength Alfv\'en waves, a
leading contender for the mechanism responsible for heating the solar
corona.  Alternatively, observing the refractive decoherence,
characterized by a sustained departure from the $\lambda^2$-law, can
place a direct limit upon the scale of the dominant wave modes.

\subsection{Pulsars: Interstellar Medium} \label{sec:PISM}
Traditionally, the observed $\RM$ of pulsars has been attributed to
the propagation through the ISM.  When coupled with a
determination of the dispersion measure, Pulsar $\RM$'s provide a
probe of the strength and distribution of the Galactic magnetic
field \cite[see, e.g.,][]{Men-Ferr-Han:08,Nout_etal:08}.  Typical
inferred magnetic field strengths for the ISM are on the order of
$1\,\muG$--$10\,\muG$, and place Pulsar polarization measurements in
the orange region labeled ``ISM'' in the lower-left of
Figure \ref{fig:regimes} for 749 objects, collected from
\citet{Nout_etal:08} and the ANTF catalog at
www.atnf.csiro.au/research/pulsar/psrcat
\citep{Manc-Hobb-Teoh-Hobb:05}.  For many pulsars, this extends above
the heliospheric cutoff, reaching in some cases $100\,\kHz$.  However,
less well defined is the refractive decoherence limit, which depends
critically upon the turbulent scale of the ISM.  That is,
$\nuR\simeq10 D_{22}^{1/3} \ell_{B,20}^{-2/3} \RM_3^{1/3}\,\kHz$, and
thus as long as the ISM varies primarily on scales above $30\,\pc$
refraction may be ignored.  On the other hand, sub-parsec turbulence in
the ISM could push $\nuR$ above $200\,\kHz$, rendering the
super-adiabatic regime inaccessible.

\subsection{X-ray Binaries} \label{sec:XB}
As the stellar mass analogs of AGN, X-ray binaries might be expected
to produce sizable $\RM$'s.  This is supported by our $\RM$ estimate
for accreting black holes.  With typical luminosities and masses on
the order of $10^{-1}\,L_{\rm Edd}$ and $10\,\Ms$,
respectively, Equation (\ref{eq:RMacc}) implies tremendous rotation
measures, $\RM\simeq8\times10^{21}\,\rad\,\m^{-2}$.  Therefore,
Faraday rotation within the accretion flow will almost certainly be
band-depolarized below the X-rays (i.e., for $\lambda \lesssim 3\,$\AA~
given a resolving power of 500). Despite this, the transition into the
super-adiabatic regime doesn't occur until roughly $40\,\eV$, making
it all but impossible to probe the accretion flow in this way.
However, the magnetic fields and densities in X-ray binary jets are
considerably lower, and there have been a handful of radio $\RM$
measurements of these features.  

At $\GHz$ frequencies, SS 433's jets are polarized at the $1\%$ level,
and have a rotation measure of roughly $-119\,\rad\,\m^{-2}$
\citep{Gilm-Seaq:80,Stir-Spec-Cawt-Para:04}.  The magnetic field strength
in the region surrounding the jets, inferred from fitting the source
structure at $5\,\GHz$,  is roughly $4\,\mG$, considerably weaker than
that associated with the core ($> 20\,\mG$) though much stronger than
the ISM field.

During outbursts, GRS 1915+105 and XTE J1748-288 have both exhibited
substantial polarizations, rising in both cases above $20\%$ level.
In GRS 1915+105, during these periods, the polarization angle of the
ejected component evolved in a fashion consistent with rotation
measures of $60\,\rad\,\m^{-2}$--$80\,\rad\,\m^{-2}$
\citep{Mill_etal:05}.  While it isn't clear what the strength of the
magnetic field is in this source, estimates based upon equipartition
arguments in the ejected jet give $B\simeq1\,\mG$.  For XTE J1748-288,
on the other hand, the rotation measure varied between
$7\,\rad\,\m^{-2}$--$12\,\rad\,\m^{-2}$, over a 10 day period
\citep{Broc_etal:07}.  During this time, the minimum energy required
to produce the radio ejection was $2.6 (d\,\kpc^{-1})^{8/7}
\times10^{42}\,\erg$, where the distance to XTE J1748-288, $d$, is
less than $8\,\kpc$.  With an volume of $9\times10^{48}\,\cm^{3}$,
inferred from the outburst timescale, this implies an equipartition
magnetic field on the order of $3\,\mG$, similar to those in GRS
1915+105 and SS 433.  

These three objects are shown in Figure \ref{fig:regimes} by the
cyan region labeled ``XB''.  The combination of relatively low $\RM$
and $\mG$ fields again places $\nuSA\simeq1\,\MHz$.  Taking a typical
scale of $10^{16}\,\cm$, this implies
$\nuR\simeq 20 D_{16}^{1/3} \ell_{B,16}^{-2/3} \RM_2^{1/3}\,\kHz$, and
therefore refraction is insignificant in XRB jets.

\subsection{Galactic Center} \label{sec:GC}
The inner $100\,\pc$ of the Milky Way is an extraordinarily active region,
exhibiting intense star formation (e.g., the Arches \& Quintuplet
clusters), numerous supernovae remnants and the mysterious non-thermal
filaments (NTF's).  Unique to the Galactic center, the NTF's are
comparatively bright, highly polarized filamentary synchrotron sources
extending in some cases for nearly $30\,\pc$.  In addition to the
large NTF's (the Southern \& Northern Threads, the Radio Arc and the
Snake), there are a number of shorter NTF's, extending for
$5,\pc$--$10\,\pc$, and in some cases misaligned with their larger
brethren \citep{Yuse-Morr:87,Lang-Morr-Eche:99,Law-Yuse-Cott:08}.

The origin of the NTF's is presently unclear.  The orientation of the
large NTF's appears to be aligned with a poloidal Galactic field,
suggesting that they are associated with magnetic flux tubes that carry
the Galactic field through the central regions.  This is supported by
the large observed polarizations, up to $70\%$ in some places, which
require the presence large-scale ordered magnetic fields
\citep[see, e.g.,][]{Yuse-Ward-Para:97}.  However, this is in conflict
with the extraordinary magnetic fields required to maintain stability
within the Galactic center's dynamic environment, nearly a $\mG$
\citep[see, e.g.,][]{Yuse-Morr:87}.  Furthermore, the presence
of smaller, misaligned filaments appears to argue against a connection
to large-scale Galactic fields \citep{LaRo-Nord-Lazi-Kass:04}.

The $\RM$ for the NTF's varies considerably along the filament, ranging
from roughly $10^2\,\rad\,\m^{-2}$ to $6\times10^3\,\rad\,\m^{-2}$
\citep{Yuse-Morr:87,Morr-Yuse:89,Gray_etal:95,Yuse-Ward-Para:97,Lang-Morr-Eche:99}.
These are quite large in comparison to the $\RM$'s of roughly
$200\,\rad\,\m^{-2}$ (implying field strengths of roughly $20\,\muG$)
observed for extragalactic sources that are viewed through the inner
$170\,\pc$ of the Galactic center \citep{Roy-Rao-Subr:08}, suggesting
that the Faraday screen must lie close to, or even are intrinsic
to, the NTF's.  This is supported by the large $\RM$ gradients
observed across some NTF's, which require turbulent $10\,\mG$ ISM
fields in the region, field strengths nearly an order of magnitude
larger than those inferred in the NTF's themselves
\citep{Yuse-Ward-Para:97}.

As a result, similar to Pulsars and AGN, there is considerable
uncertainty regarding the region responsible for the observed Faraday
rotation, and therefore the relevant magnetic field strength.
Furthermore, given the dynamic nature of the Galactic center region it
is additionally unclear if the field is ordered or turbulent.  The
red region labeled ``GC'' in the middle of Figure \ref{fig:regimes} shows
the best guess values for the NTF's the Northern Thread, the Snake,
the Radio Arc and G 359.54+0.18, and roughly characterizes the
uncertainty in the magnetic field.  By comparison, taking a fiducial
NTF transverse scale of $0.1\,\pc$ and line-of-sight length of $3\,\pc$
gives a typical $\nuR\simeq40\,\kHz$, well below $\nuSA$.  This is
only weakly dependent upon the location of the Galactic center Faraday
screen, rising to $0.1\,\MHz$ for distances of up to $100\,\pc$.

\subsection{Galactic Molecular Clouds} \label{sec:GMC}
Magnetic turbulence is thought to be critical for supporting
structures within molecular clouds and generating the dense molecular
cores that are the sites of star formation \citep{McKe-Ostr:07}.
Despite this, there is still considerable uncertainty regarding the
magnetic field strengths and geometry within individual objects, and
therefore the role that magnetic fields play in the process of star
formation.  Typical values of the magnetic field in Molecular clouds
range from Galactic value of a few $\muG$ to nearly a $\mG$ inferred
from Zeeman splitting of molecular lines \
\citep[see, e.g.][]{Crut:99,Han-Zhan:07}. 
The measured molecular lines are presumably produced in the densest
regions, and therefore indicative of the fields nearest the sites of
star formation.  Nevertheless, the comparative lack of large magnetic
fields in stars implies that at some point during the star formation
process the magnetic fields must somehow dissipate (either by
diffusion or reconnection), and thus these may not represent the
strongest magnetic fields present in molecular clouds.

High-resolution maps of the rotation measures of molecular clouds have
been obtained via $21\,\cm$ observations of nearby star forming
regions \citep[see, e.g.,][]{Uyan_etal:03,Woll-Reic:04}.  These
generally find considerable substructure and exhibit large $\RM$
gradients.  The average average rotation measures are on the order of
$30\,\rad\,\m^2$, though with a dispersion of roughly
$3\times10^2\,\rad\,\m^2$ about the mean value, and reaching
$10^3\,\rad\,\m^{-2}$ in some places.  This puts molecular clouds in
the lower-left of Figure \ref{fig:regimes}, denoted by the dark red
region and labeled ``MC'', and $\nuSA\simeq30\,\kHz$--$2\,\MHz$,
reflecting the uncertainty in the underlying physical conditions.

As with other sources, the value of $\nuSA$ would all but determine
the location of the Faraday rotating medium.  Furthermore, comparing
$\RMSA$ and $\RMA$ would provide an independent measurement of the
magnetohydrodynamic turbulence believed to support these clouds.  In
particular, this would allow a comparison between the geometry of the
magnetic field and that expected from the turbulent velocity field.

As with other sources, uncertainty arises in $\nuR$ associated with
the uncertain scales of the location of and variations in the Faraday
screen.  Nevertheless, even for large molecular clouds ($10^2\,\pc$)
and relatively small turbulent structures ($0.1\,\pc$) the refractive
decoherence frequency is still only roughly $60\,\kHz$, almost
certainly well below the relevant estimates for $\nuSA$.

\subsection{Starburst Galaxies} \label{sec:SG}
Infrared galaxies are presumed to be undergoing a phase of prodigious
star formation.  Therefore, one might expect that these offer the
prospect of measuring the properties of extragalactic star forming
regions generally, and extragalactic molecular clouds specifically.
Radio observations of infrared galaxies have indeed shown typical
$\RM$'s of $10\,\rad\,\m^{-2}$, though with a dispersion of roughly
$50\,\rad\,\m^{-2}$ (though some reach nearly $200\,\rad\,\m^{-2}$),
these are somewhat more narrowly distribute about the mean
\citep{Heal-Brau-Edmo:09}.  This is a reflection of the fact that we are
necessarily averaging over a much larger region than we can resolve in
Galactic molecular clouds.  Indeed, the $\RM$ for the individual
galaxies is comparable to that averaged over individual Galactic
molecular clouds \citep{Uyan_etal:03,Woll-Reic:04}.

Again there is considerable uncertainty in the strength of the
magnetic field in the star forming regions of infrared galaxies.  We
adopt a typical value of $10\,\muG$, though the dynamics of recent
galactic interactions may raise this substantially.  Nevertheless, the
region occupied in Figure \ref{fig:regimes} is between molecular clouds
and the ISM, labeled ``IG'' and shown in magenta.  This puts
$\nuSA\simeq10^2\,\kHz$.  This is, however, becoming uncomfortably
close to the refractive decoherence scale, comparable to the
$60\,\kHz$ obtained in section \ref{sec:GMC}.  Thus, in this case even
detection of the super-adiabatic regime will shed light upon the
conditions within the star factories of starburst galaxies.

\subsection{Intracluster Medium} \label{sec:ICM}
The magnetic field embedded in the hot, turbulent gas at the centers
of galaxy clusters has an estimated strength of
$10\,\muG$--$30\,\muG$.  Combined with the $\kpc$ path lengths, dense
cluster cores are capable of produce substantial $\RM$'s.

Indeed, the fact that \HzRG's are located at the centers of
cooling-core clusters suggests that the anomalously large $\RM$'s
observed in these sources may be associated with the intracluster
medium (ICM) and are not intrinsic to the galactic nuclei.  This
produces estimates for the magnetic field 1--2 orders of magnitude
lower than those obtained in Section \ref{sec:UAGN}, roughly on the
order of $20\,\muG$.  These are shown in the dark-orange region labeled
``\HzRG-c'', implying that in the {\em rest-frame}
$\nuSA\simeq1\,\MHz$ for these galaxies.  Considering the high
redshifts of these sources, this gives $\nuSA\simeq300\,\kHz$.
This is well above typical estimates of $\nuR$, which for a $20\,\kpc$
core exhibiting large-scale structure on $100\,\pc$ scales is still
roughly only $20\,\kHz$.

Observations of typical radio galaxies embedded in the centers of clusters
find rotation measures as high as $3\times10^3\,\rad\,\m^{-2}$
\citep{Ge-Owen:93,Ge-Owen:94}.  Representative of these is J1346+268,
located near the center of Abell 1795.  Combined with electron density 
estimates from X-ray observations, this gives a lower-limit upon the
cluster magnetic field strength of $20\,\muG$.  This is roughly an
order of magnitude lower than the ICM equipartition value, suggesting
that the magnetic field is likely to be tangled on scales small in
comparison to the cluster size.  Producing the observed $\RM$ with a
stochastic field requires roughly $N=10^2$, resulting in an order of
magnitude difference in $\RMSA$ and $\RMA$.  However, the
corresponding $\nuSA$ is roughly $300\,\kHz$, comparable to that found
for the \HzRG, assuming that the Faraday rotation occurs within the
ICM in those sources.

\section{Conclusions} \label{sec:C}
Despite commonly being described in terms of the independent
propagation of plasma modes, non-adiabatic effects at magnetic field
reversals force the mode crossings that are critical to understanding
the standard expressions for Faraday rotation.  For nearly all
astrophysical systems, at frequencies above a few $\GHz$, these
effects dominate the propagation at magnetic field reversals.
However, at sufficiently low frequencies it is possible to enter a
``super-adiabatic'' regime, in which the two plasma modes propagate
independently throughout, resulting in rotation measures that are
proportional to $\int n_e |B_\parallel| \d\ell$ instead of
$\int n_e B_\parallel \d\ell$.

For large-scale ordered magnetic fields, there is little difference
between the standard and super-adiabatic regimes.  However, for
magnetic field geometries exhibiting many reversals along the line of
sight, the rotation measure in the super-adiabatic regime can be
substantially larger than that determined at high frequencies.  Thus,
comparisons between the standard and super-adiabatic rotation provides
a unique means to measure the magnetic field geometry along the line
of sight.

The frequency at which Faraday rotation transitions between the standard and
super-adiabatic regimes, $\nuSA$, is itself dependent upon the local plasma
properties at the magnetic field reversals, though only weakly dependent
upon the magnetic field geometry.  Thus even the detection
of super-adiabatic Faraday rotation, marked by a rapid increase in the
rotation measure at low frequencies, provides localized information
about the plasma density and magnetic fields.  This has the potential
to remove the degeneracy between propagation path length and plasma
density or magnetic field strength, allowing an unambiguous
determination of the location of the Faraday screen in many systems.

For known systems, $\nuSA$ ranges from the heliospheric cutoff,
approximately $10\,\kHz$, to nearly $10\,\GHz$, with regions
containing stronger magnetic fields and higher rotation measures
transitioning at higher frequencies.  For resolved active galactic
nuclei, such as Sgr A*, M81* and M31, this lies above the ionospheric
cutoff at $15\,\GHz$, and hence is observable from ground based radio
telescopes.  Some unresolved AGN, \HzRG~and Pulsars may be observable from
the ground as well.  Detection of super-adiabatic Faraday rotation in
any of these sources would provide striking confirmation of the
presence of in situ Faraday rotation.
For X-ray binaries, infrared galaxies, molecular clouds, the Galactic
center region, intracluster medium, interstellar medium and the solar
corona, the super-adiabatic regime should be observable from
space-based radio observatories.  

The search for super-adiabatic Faraday rotation is especially timely
given the proliferation of ultra-low frequency telescopes, such as the
{\em Murchison Widefield Array} in Australia
($80\,\GHz$--$300\,\GHz$) and the {\em Low Frequency Array} in Europe
($10\,\MHz$--$240\,\MHz$), built ostensibly for the purpose of mapping
$21\,\cm$ emission from the high-$z$ Universe, and the
{\em Frequency Agile Solar Radiotelescope} in North America
($50\,\MHz$--$21\,\GHz$), currently under development \citep{Bast:03}.
All of these devices will collect the full
compliment of Stokes parameters, and thus will be capable of searching
for super-adiabatic Faraday rotation.  In the far term, space-based
radio observatories, such as {\em Astronomical Low Frequency Array},
have been discussed for more than two decades now \citep{Jone_etal:00}.
They provide the only means to access the $10\,\kHz$--$15\,\MHz$ radio
window, and beyond the confines of the Earth, can in principle do so
with resolutions comparable to much shorter wavelengths at the
{\em Very Long Baseline Array}.  The ability to directly probe the
geometry of astrophysical magnetic fields along the line of sight
should add to the list of science motivators for such a capability.

\section*{Acknowledgements}
This research has made use of data from the MOJAVE database that is
maintained by the MOJAVE team (Lister et al., 2009, AJ, 137, 3718).
This work was supported in part by the US Department of Energy under
contract number DE-AC02-76SF00515.

\appendix
\section{Radiative Transfer Regimes} \label{RTR_App}
In general, Maxwell's equations give
\begin{equation}
\left( \nabla^2 - \bmath{\nabla} \bmath{\nabla} + \omega^2
\bmath{\epsilon} \right) \cdot {\bf E} = 0 \,,
\end{equation}
for an electric field ${\bf E}$, and a dielectric tensor
$\bmath{\epsilon}$.  For plane waves propagating along the
$z$-axis in a plane parallel medium (we address the possibility of
refraction in the following appendix), this reduces to
\begin{equation}
\frac{\d^2 {\bf F}}{\d z^2} + \omega^2 \bmath{\epsilon} \cdot {\bf F}
= 0\,,
\label{plane_wave}
\end{equation}
where ${\bf F}$ is the Jones vector (\ie,~a two-dimensional vector
constructed from the transverse components of ${\bf E}$).  For an
anisotropic dielectric tensor, there will exist two nondegenerate
transverse modes defined such that
\begin{equation}
\omega^2 \bmath{\epsilon} {\bf F}_i = k_i^2 {\bf F}_i\,.
\end{equation}
In the case of a plasma, these are in general elliptically polarized
and aligned with the component of the background magnetic field normal to the
direction of propagation, \ie,
\begin{equation}
{\bf F}_1 = {\bf Q}
\left(\begin{array}{c}
\sin\chi\\
i\cos\chi
\end{array}\right)
\quad{\rm and}\quad
{\bf F}_2 = {\bf Q}
\left(\begin{array}{c}
\cos\chi\\
-i\sin\chi
\end{array}\right)\,,
\end{equation}
where the orientation of the polarization ellipses with respect to a
set of axis fixed in space is determined by
\begin{equation}
{\bf Q} =
\left(\begin{array}{cc}
\cos\phi &\sin\phi\\
-\sin\phi &\cos\phi
\end{array}\right)\,.
\end{equation}
Then, ${\bf F} = F_1 {\bf F}_1 + F_2 {\bf F}_2$, may be inserted into Equation
(\ref{plane_wave}) to give
\begin{align}
F_1''
+&
2 i s_2 \varphi F_1'
+
\left(k_1^2-\varphi^2-\psi^2+is_2\varphi'+2ic_2\varphi\psi\right) F_1
\nonumber \\
&=
2\left(\psi-ic_2\varphi\right)F_2'
+\left(\psi'-ic_2\varphi'+2is_2\varphi\psi\right)F_2
\nonumber \\
F_2''
-&
2 i s_2 \varphi F_2'
+
\left(k_2^2-\varphi^2-\psi^2-is_2\varphi'-2ic_2\varphi\psi\right)F_2
\nonumber \\
&=
-2\left(\psi+ic_2\varphi\right)F_1'
-\left(\psi'+ic_2\varphi'-2is_2\varphi\psi\right)F_1 \,,
\label{forsterling}
\end{align}
where a prime denotes differentiation with respect to $z$,
$c_2\equiv\cos2\chi$, $s_2\equiv\sin2\chi$, $\varphi\equiv\phi'$ and
$\psi\equiv\chi'$.  When $\varphi=0$, these reproduce F\"orsterling's
coupled equations \citep[\cf][]{Budd:61,Ginz:70}.

Thus far, no approximations have been made regarding the wave length
or scale lengths of the plasma.  From the form of Equations
(\ref{forsterling}), it is clear that if $\phi$ and $\psi$ vanish, the
two modes will propagate completely independently.  In the limit that
$\varphi$ and $\psi$ are small in comparison to $k_{1,2}$, we may look
for WKB solutions of the form
\begin{equation}
F_i = \frac{f_i}{\sqrt{k_i}}\,\e^{i\int k_i \d z} \,,
\end{equation}
and hence
\begin{align}
f_1' + i s_2 \varphi f_1 &=
\left(\psi-ic_2\varphi\right)f_2\,\e^{-i\int\Delta k \d z}\nonumber\\
f_2' - i s_2 \varphi f_2 &=
-\left(\psi+ic_2\varphi\right)f_1\,\e^{i\int\Delta k \d z}\,,
\end{align}
where $\Delta k\equiv k_1-k_2$ terms on the order of $\psi^2$,
$\varphi^2$, $\psi'$, $\varphi'$, $\psi\varphi$, and $f_i''$ were
ignored as they are small by assumption relative to those that remain.
Further expand the $f_i$ as
\begin{equation}
f_1 = u_1 \e^{-i\int(\Delta k/2) \d z}
\quad{\rm and}\quad
f_2 = u_2 \e^{i\int(\Delta k/2) \d z}\,.
\end{equation}
Then,
\begin{align}
u_1' - i \left(\frac{\Delta k}{2} - s_2 \varphi\right) u_1
&=
\left(\psi - i c_2 \varphi\right) u_2\nonumber\\
u_2' + i \left(\frac{\Delta k}{2} - s_2 \varphi\right) u_2
&=
-\left(\psi + i c_2 \varphi\right) u_1\,,
\end{align}
which may be combined to give
\begin{align}
&u_1'' + \left[ \left(\frac{\Delta k}{2}\right)^2
+ \psi^2 + \varphi^2 - s_2 \varphi \Delta k - i\frac{\Delta k'}{2}
\right] u_1 = 0\nonumber\\
&u_2'' + \left[ \left(\frac{\Delta k}{2}\right)^2
+ \psi^2 + \varphi^2 - s_2 \varphi \Delta k + i\frac{\Delta k'}{2}
\right] u_2 = 0 \,.
\label{u12_eqs}
\end{align}

If the $\psi$ and $\varphi$ terms are dominated by the $\Delta k$
terms, then
\begin{equation}
u_{1,2} \simeq {\rm const} \times
\e^{\pm i \int \left(\Delta k/2\right) \d z} \,,
\end{equation}
and thus the $f_i$ are constant.  Therefore, in this limit the modes
propagate independently (the so-called adiabatic regime).  In the
opposing limit, when $\psi$ and $\varphi$ dominate $\Delta k$, then
Equations (\ref{u12_eqs}) are indistinguishable from the isotropic
case (\ie,~$\Delta k = 0$), and therefore the polarization propagates
unaltered (the so-called vacuum regime).  This can be
directly proved by solving for $u_{1,2}$ in this limit and expressing
the answer in terms of ${\bf F}$.  Because the $\varphi \Delta k$ term
will only be relevant when
\begin{equation}
\left( \frac{\Delta k}{2} \right)\varphi \sim \psi^2 + \varphi^2 \,,
\end{equation}
this term may be neglected in determining the limiting regimes,
yielding
\begin{equation}
\begin{array}{lcl}
\displaystyle
\psi^2+\varphi^2 \ll \left(\frac{\Delta k}{2}\right)^2 + \left|
\frac{\Delta k'}{2}\right|
&\quad&{\rm adiabatic}\\
&&\\
\displaystyle
\psi^2+\varphi^2 \gg \left(\frac{\Delta k}{2}\right)^2 + \left|
\frac{\Delta k'}{2}\right|
&\quad&{\rm vacuum.}
\end{array}
\label{eq:av_cond}
\end{equation}

What remains is to identify these limits in terms of the physical
characteristics of the plasma.  That $\nu$ exceeds the upper cutoff of
the extraordinary mode requires $X+Y<1$, which in turn requires that
the propagation occur above both the electron cyclotron and electron
plasma frequencies.  Under these circumstances the ionic contribution
to the dielectric tensor may be safely ignored.  Thus, the $k_{1,2}$ are
given by the standard Appleton-Hartree dispersion relations:
\begin{equation}
k_{1,2} = \omega \sqrt{ 1 - \frac{2X(1-X)}{2(1-X)-Y^2\sin^2\theta \mp \Gamma}}
\end{equation}
where
\begin{equation}
\Gamma = \sqrt{ Y^4 \sin^4\theta + 4 Y^2(1-X)^2\cos^2\theta}\,.
\end{equation}
To second order in $Y$, this gives
\begin{equation}
\Delta k
\simeq
\frac{\omega}{c} X Y
\sqrt{ \cos^2\theta + \left(\frac{Y}{2}\right)^2\sin^2\theta }
\,.
\label{Delta_k_0}
\end{equation}
The ellipticity angle is given by \citep[see \eg][]{Ginz:70,Budd:64}
\begin{equation}
\tan \chi = x + \sgn(x) \sqrt{1 + x^2}
\end{equation}
where
\begin{equation}
x \equiv \frac{Y \sin^2\theta}{2 \left(1 - X \right)\cos\theta} \,.
\label{x_from_mu}
\end{equation}
As a direct result,
\begin{equation}
\chi'
=
\frac{x'}{2(1+x^2)}
\simeq
\frac{Y}{4} \frac{(1+\cos^2\theta)\sin\theta}{(Y/2)^2 + \cos^2\theta}
\theta'\,,
\end{equation}
where the finally expression assumes that $Y \ll 1$ and
$X',Y'\ll\theta'$.  Note that for $\theta\simeq\pi/2$,
$\chi' \simeq Y^{-1} \theta' \simeq Y^{-1} \phi'$ which at high
frequencies ($Y\ll 1$) is generally much larger than and
$\phi'$, and thus $\psi$ dominates $\varphi$ near magnetic field
reversals in the left-hand sides of eqs. (\ref{eq:av_cond}), and in
particular is maximized at $\theta=\pi/2$.

On the other hand, as long as $X',Y'\ll\phi$ near $\theta=\pi/2$, $|\Delta k'/2|$
vanishes identically, due to the symmetry of $\Delta k$.  This does
not mean that $|\Delta k'/2|$ doesn't dominate $(\Delta k/2)^2$
elsewhere, which is clearly the case for the parameters shown in
Figure \ref{fig:FR}.  Nevertheless, sufficiently close to
$\theta=\pi/2$, the right-hand sides of eqs. (\ref{eq:av_cond}) are
dominated by the $(\Delta k/2)^2$ term, and reach their minimum.

Therefore, for the mode propagation to remain adiabatic for all
magnetic field orientations (i.e., through an entire magnetic field
reversal), it is both necessary and sufficient to have
\begin{equation}
\psi \ll \left|\frac{\Delta k}{2}\right|\,.
\end{equation}
It is this condition that we use to define the ``super-adiabatic'' regime.
Note that at magnetic reversals
\begin{equation}
\left| \frac{\Delta k}{2} \right| \phi
\ll
\left| \frac{\Delta k}{2} \right| \psi
<
\max\left[\psi^2,\left(\frac{\Delta k}{2}\right)^2\right]\,,
\end{equation}
and therefore, much smaller than the dominant term in the above
adiabatic condition, either $\psi^2$ or $(\Delta k/2)^2$, as was previously claimed.

\section{Refraction and Faraday Decoherence} \label{sec:refract}
In the previous section we ignored the refraction of the radio wave.
For propagation near the plasma and/or cyclotron frequencies this is
not justified.  Therefore, here we address the corrections and limits that
refraction places upon the computation of the Faraday rotation.

Refraction potentially enters into the problem in two ways.  First
it provides a third scale over which the plasma eigenmodes change,
though typically larger than $2\pi/\chi'$ and $2\pi/\phi'$, and thus
not relevant.  Second, and much more important, is the possibility
that the two plasma eigenmodes refract differently, propagating
through significantly different regions and sampling distinct plasma
conditions.  It is this latter effect that we concern ourselves.

Let us define $\bmath{x}_{1,2}(\eta)$ and $\bmath{k}_{1,2}(\eta)$ be
the trajectories and wave-vectors of the two plasma eigenmodes,
parametrized such that at $\eta=0$ they begin at the source and at
$\eta=1$ they arrive at the observer.  The phase difference between
the two modes is then
\begin{equation}
\int_0^1d\eta \left(
\bmath{k}_1\cdot\frac{d\bmath{x}_1}{d\eta}
-
\bmath{k}_2\cdot\frac{d\bmath{x}_2}{d\eta}
\right)\,.
\end{equation}
That this may be reduced to an integral along a single curve, e.g.,
$\bmath{x}=(\bmath{x}_1+\bmath{x}_2)/2$, requires that the two curves
never separate further than the local plasma correlation length,
roughly $\ell_B$, leading to refractive decoherence.  Note that in
this case, it is not the total diffraction of the trajectories, but
the differential trajectory that matters.  Generally, this is a
complicated global condition, requiring the complete construction of
$\bmath{x}_{1,2}$ prior to addressing.  However, in the
``thin-screen'' approximation, we can obtain a simple condition for
when refractive decoherence may be safely ignored.

\begin{figure}
\begin{center}
\includegraphics[width=\columnwidth]{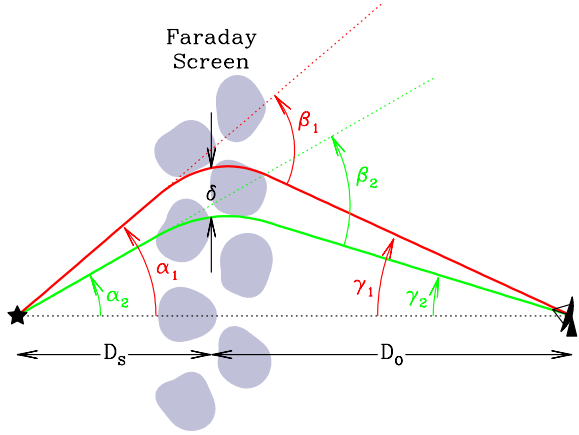}
\end{center}
\caption{Diagram of the relevant angles and distances within the
  thin-screen approximation for the fast (green) and slow (red) modes.
  The maximum deviation is $\delta$, located within the screen
  itself.} \label{fig:lens}
\end{figure}

Within the thin-screen approximation we assume that the Faraday
screen's width is much smaller than both, the distance between the source
and the screen, $D_S$, and the distance between the screen and the
observer, $D_O$.  Within this approximation, the trajectories are
roughly straight lines, joined by a lensing angle at the screen.
Geometrically, it is easy to see that $D_S\alpha_i = D_O\gamma_i$ and 
$\alpha_i+\gamma_i = \beta_i$, from which it follows directly that,
\begin{equation}
\delta = \frac{D_S D_O}{D_S+D_O} \left(\beta_1-\beta_2\right)\,.
\end{equation}
That the maximum deviation between the paths taken by the two plasma
eigenmodes is less than the characteristic plasma
correlation length reduces to $\delta\ll\ell_B$, or
\begin{equation}
|\beta_1-\beta_2|
\ll
\frac{D_S+D_O}{D_S D_O} \ell_B
= \frac{\ell_B}{D}
\,,
\end{equation}
where $D^{-1}\equiv D_S^{-1} + D_O^{-1}$.  At this point, we must
determine $\beta_i$.

It suffices to consider propagation in the quasi-longitudinal regime
since this dominates the differential refraction (because the
difference in the indices of refraction is largest and the
propagation nearly always occurs in this regime).  Therefore, the
dispersion relation is given by
\begin{equation}
D_{1,2} \simeq
\frac{1}{2}\left[k_{1,2}^2 - \omega^2 + \omega^2 X(1\pm Y) \right]\,,
\end{equation}
i.e., $D_{1,2}$ vanishes along the ray.  The equations defining the
ray are
\begin{equation}
\begin{array}{c}
\displaystyle
\frac{d\bmath{x}_{1,2}}{d\eta}
=
\frac{\partial D_{1,2}}{\partial\bmath{k}_{1,2}}
=
\bmath{k}_{1,2}\,,\\
\\
\displaystyle
\frac{d\bmath{k}_{1,2}}{d\eta}
=
-\frac{\partial D_{1,2}}{\partial\bmath{x}_{1,2}}
=
-\frac{\omega^2}{2}\frac{\partial X(1\pm Y)}{\partial\bmath{x}_{1,2}}\,.
\end{array}
\end{equation}
Outside of the Faraday screen, $X=Y=0$, and $k_i^2=\omega^2$.  In the
weak deflection limit, we may treat the component of $\bmath{k}_i$
along the original direction as unchanged, the refraction simply
introducing an orthogonal component.  That is, we set
\begin{equation}
\beta_{1,2}
\simeq
-
\omega^{-1} \int d\eta
\frac{d\bmath{x}_{1,2}}{d\eta}
\times
\frac{\partial X(1\pm Y)}{\partial\bmath{x}_{1,2}}
\simeq
-\frac{1}{2} X(1\pm Y) f
\end{equation}
where $f$ is the characteristic fractional variation in the plasma
parameters on the scale of $\ell_B$.  Therefore, for a Faraday screen
of width $\ell_B$ we obtain $|\beta_1-\beta_2|\simeq XYf$.  Should the
screen be composed of may turbulent cells, the trajectories will
diffuse, with
$|\beta_1-\beta_2|\simeq\sqrt{N}XYf\simeq\sqrt{L/\ell_B}XYf$.
Therefore, for the maximum deviation between the paths of the two plasma
eigenmodes to be sufficiently small we require
\begin{equation}
XYf \ll \frac{\ell_B^{3/2}}{D L^{1/2}}\,.
\end{equation}
As with the super-adiabatic condition, we may view this physically, as
a lower limit upon $\ell_B$, or observationally, as an upper limit
upon $\nu$.  In terms of the latter, we have
\begin{equation}
\nu \gg
\left(\nu_P^2 \nu_B f \frac{D L^{1/2}}{\ell_B^{3/2}}\right)^{1/3}\,,
\end{equation}
the expression underlying those in Equation (\ref{eq:refrac}).


\end{document}